# Submillimeter and Far-InfraRed Experiment (SAFIRE): A PI class instrument for SOFIA

R. A. Shafer[a], S. H. Moseley, Jr.[a], P. A. Ade[b], D. J. Benford[a],
G. Bjoraker[a], E. Dwek[a], D. A. Neufeld[c], F. Pajot[d], T. G. Phillips[e], G. J. Stacey[f]

[a] NASA Goddard Space Flight Center, Code 685, Greenbelt, MD 20771
[b] Queen Mary and Westfield College, Mile End Road, London E14NS, UK
[c] Johns Hopkins University, 3400 N. Charles St., Baltimore, MD 21218
[d] IAS, Bât. 121, Université Paris Sud, 91405, Orsay, France
[e] California Institute of Technology, Mail Code 320-47, Pasadena, CA 91125
[f] Cornell University, Space Sciences Building, Ithaca, NY 14853

**ABSTRACT**

SAFIRE is a versatile imaging Fabry-Perot spectrograph covering 145 to 655 microns, with spectral resolving powers ranging over 5-10,000. Selected as a "PI" instrument for the airborne Stratospheric Observatory for Infrared Astronomy (SOFIA), SAFIRE will apply two-dimensional pop-up bolometer arrays to provide background-limited imaging spectrometry. Superconducting transition edge bolometers and SQUID multiplexers are being developed for these detectors. SAFIRE is expected to be a "First Light" instrument, useable during the initial SOFIA operations. Although a PI instrument rather than a "Facility Class" science instrument, it will be highly integrated with the standard SOFIA planning, observation, and data analysis tools.

## 1. INTRODUCTION

The submillimeter is the last major window in the electromagnetic spectrum to be opened. A wealth of lines of atoms and molecules which are diagnostics of all the major components of the interstellar medium and star forming regions are found in the submillimeter (Fig. 1). In spite of this, the scientific exploitation of this spectral region has progressed slowly, because the required technologies for high performance instruments were not available. We describe here the Submillimeter and Far InfraRed Experiment (SAFIRE) for SOFIA. SAFIRE is a versatile imaging Fabry-Perot spectrograph covering the spectral region from 145 to 655 μm with spectral resolving powers ranging from 5 to $10^4$. SAFIRE can determine the energy balance and physical conditions in many important phases of the interstellar and circumstellar environment by imaging these regions in lines arising from molecular, atomic and ionized components, along with the dust continuum emission. These observations are critical to achieving a clear understanding of the processes which result in the formation of stars, and which control star formation on a galactic scale in starburst galaxies. SAFIRE can also examine active galactic nuclei, cooling flows in galaxy clusters, circumstellar debris disks, and the chemistry of the outer planets.

SAFIRE's spectral region includes lines tracing all the major components of gas in astrophysical environments, ranging from lines of hydrides which trace the densest, coldest components of molecular clouds, to excited rotational transitions of heavier molecules, which trace the warmer regions of molecular clouds over a range of densities and temperatures. These heavier molecules (CO, for example) allow us to trace warm gas in cloud cores, heated by a variety of equilibrium processes, and hot molecular gas heated by the passage of nondissociative shocks into the cloud. Several atomic transitions, such as CI (370 and 609μm), CII (158μm), and NII (205μm), are important and ubiquitous tracers of the energetics of the ISM. Finally, SAFIRE will provide imaging in the dust continuum emission and long slit grating spectroscopy to determine overall spectral energy distributions over this spectral range, providing accurate estimates of emitting dust mass and energy density in clouds.

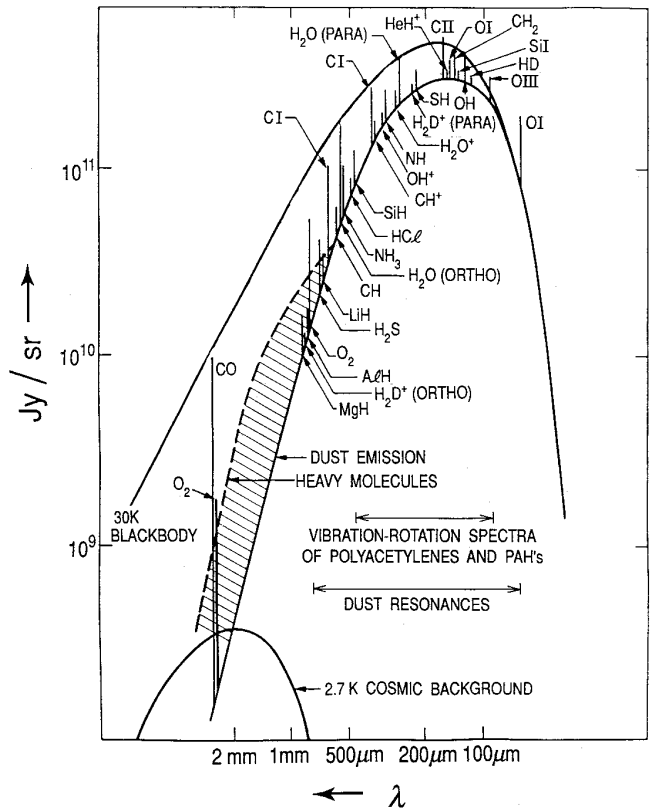

**Figure 1.** A schematic presentation of some lines in the submillimeter and FIR bands, expected in the spectrum of molecular clouds[1].

The instrument uses two 6 x 32 filled arrays of bolometers to provide background-limited imaging over a 5.7′ x 1.1′ field of view (11″ pixels with spectral resolutions ranging from 5 to $10^4$. For many investigations, this instrument approaches the fundamental limits of sensitivity possible on the SOFIA platform, and it offers a spatial multiplex advantage of a factor of 200 - 1000 over single detectors. This capability is enabled by the development over the last 15 years of sensitive bolometer arrays operating at T < 0.1 K, and is enhanced by the development of Pop-Up Detectors (PUDs) at GSFC. The PUD architecture for bolometers allows an ideal intimate connection of detectors and amplifiers in close-packed two dimensional arrays, in the same spirit as the large two dimensional arrays that revolutionized near- and mid-infrared astronomy. This results in a much more robust detector system than can be achieved using remote amplifiers. Coverage of the entire 145 – 655 μm range is possible using a broad band Fabry-Perot reflector recently developed by the SAFIRE team. A grating mode has been included for long-slit spectroscopy over the same spectral range with resolving power of ~100.

Superconducting transition edge bolometers and SQUID multiplexers are being developed by the NIST/GSFC collaboration for SAFIRE and other future missions[2]. In addition to being a robust, low-noise amplifier, the SQUID multiplexing allows for a substantial reduction in the wiring needed for the bolometer array. The current program should produce 6 x 32 multiplexed arrays using these technologies during the coming year. We will install an advanced 32 x 32 array of detectors in SAFIRE as the technology improves.

## 2. INSTRUMENT DESIGN DESCRIPTION

The SAFIRE instrument is a versatile imaging spectrometer covering the region from 145 to 655 μm. with resolving powers ranging from 5 to $10^4$. Figs. 2 & 3 show the instrument configuration and layout. The SAFIRE team has made extensive use of existing hardware and designs to reduce cost and development risk. SAFIRE consists of a dewar to provide a cryogenic environment and the mounting structure for the instrument on the SOFIA telescope; an optical system that includes six mechanisms and their optics (including a grating) to provide diffraction limited spatial resolution; two close-packed 2-dimensional arrays of silicon bolometers to provide background limited sensitivity and large spectral or spatial coverage; and three electronic subsystems to provide instrument control and data processing.

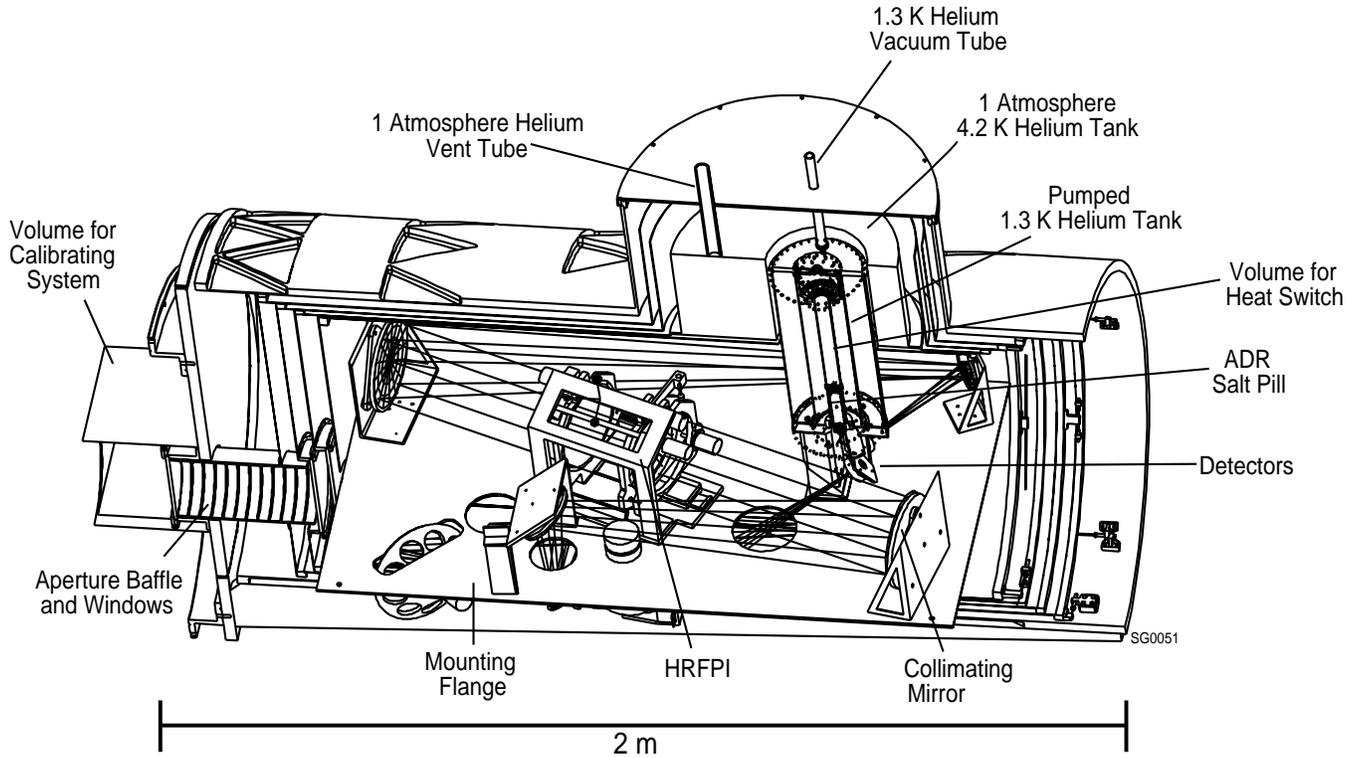

**Figure 2**. SAFIRE upper optics deck inside dewar (cutaway).

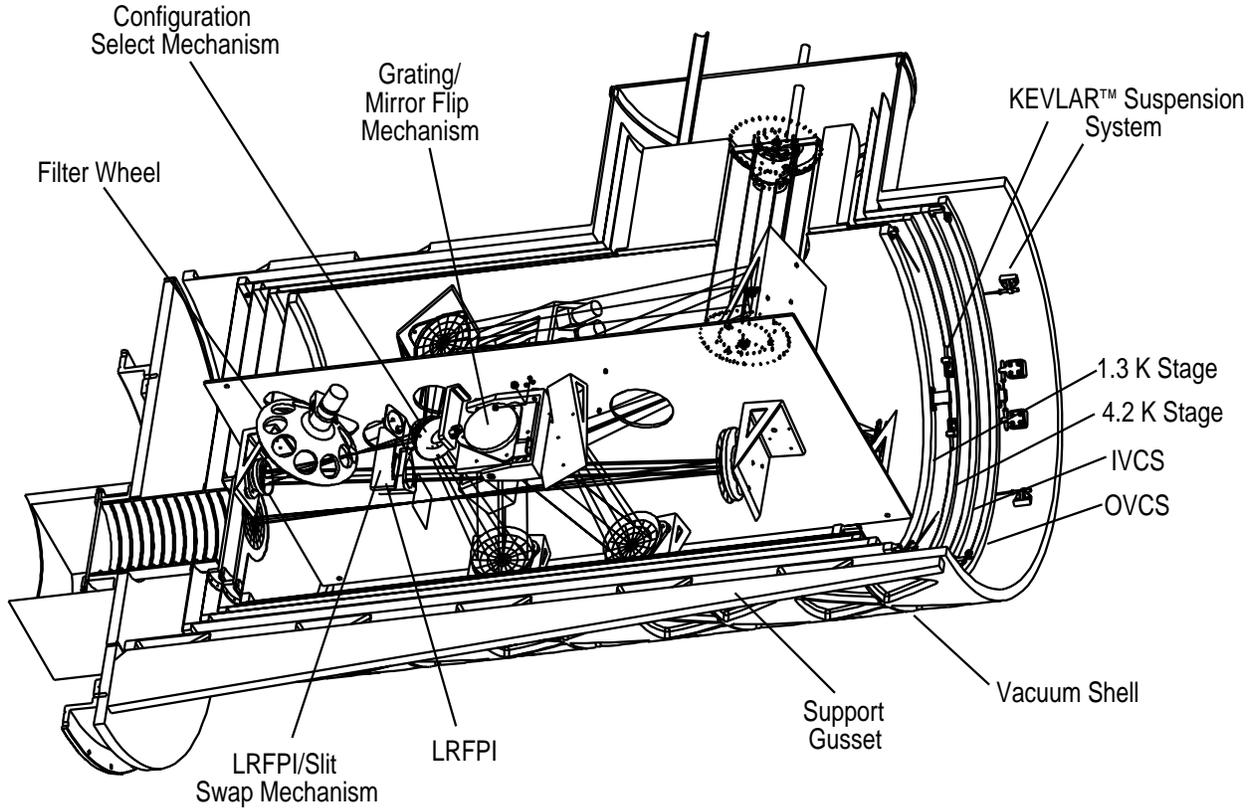

**Figure 3.** SAFIRE lower optics deck inside dewar (cutaway).

The SAFIRE instrument switches among 3 optical configurations: high (R ~1,000 to 10,000) and low (R ~100) spectral resolution imaging configurations, and a low (R ~100) spectral resolution grating spectroscopy configuration. The high resolution imaging configuration uses a large, High Resolution Fabry-Perot Interferometer (HRFPI) as a tunable bandpass filter, with a smaller, Low Resolution FPI (LRFPI) acting as the order sorter. In this mode, the beam finally impinges on the Low Background Detector Array (LBDA). In the low resolution imaging configuration, after passing through the LRFPI, the beam is diverted and focused onto the second detector, the High Background Detector Array (HBDA). The two separate detectors are required to provide adequate sensitivity over the full range of spectral resolutions, hence background powers. In the grating configuration, the LRFPI is removed from the beam and replaced by a slit, and the fold mirror, which has the grating on the back side, is flipped over. The spectrum is then projected onto the HBDA. A filter wheel with various bandpass filters completes the order sorting for the various configurations. For calibration, a flip mirror mounted in the Nasmyth tube intercepts the light from the telescope and directs a beam from the Gunn Diode/ harmonic multiplier tunable calibration source into the instrument. During the initial phase of SAFIRE development, the HRFPI mode will be available; subsequent improvements will add the LRFPI and grating spectroscopy modes.

SAFIRE uses all reflective optical elements. All optics and internal mechanisms are cooled down to 1.3 K using pumped liquid Helium. The two detectors use the GSFC-developed Pop-Up Detector (PUD) 2-dimensional bolometer arrays with superconducting transition edge thermistors (§4). These are cooled to 0.065 K using an Adiabatic Demagnetization Refrigerator (ADR). The ADR and detectors are designed to be removable from the main cryostat as a unit, with all wiring intact, for development and checkout in a small, separate dewar. An atmospheric pressure Helium stage at 4.2 K surrounds the 1.3 K stage. The boiloff vapor from the 4.2 K is used to cool inner and outer radiation shields. Everything is then housed in an aluminum vacuum shell, which forms the structural attachment to the telescope.

The optical design provides diffraction limited spatial resolution, a maximum spectral resolution of 10,000, and various spectral resolution and image formats derived from the science requirements. Table 1 summarizes the optical performance requirements and their design implementation.

**Table 1.** Optics Requirements and design implementation enable SAFIRE science.

| Derived Requirement | Key Design Parameters |
|---|---|
| Diffraction limited spatial resolution, 145 to 655 μm | • f/7.8 image on 1 mm pixels<br>• 10.6 arcsec per pixel |
| 6 x 32 detector array (upgrade to 32 x 32) | • 5.7 x 1.1 arcmin field<br>  (upgrade to 5.7 arcmin square) |
| Spectral resolution up to R = 10,000 | • 200 mm pupil diameter<br>• Up to 65 mm FPI mesh separation<br>• LRFPI and filters for order sorting |
| Intermediate field and pupil images | • f/5.7 intermediate field image<br>• 38, 100, and 200 mm diameter pupil images |
| Background limited performance | • 1.3 K dewar environment<br>• 0.065 K detector stage and baffles |
| Interface to SOFIA | • 8 arcmin diameter field, 450 mm from flange<br>• f/19.6 entrance pupil, 7.3 m ahead of focus |

Other features of the optical design provide reliability, robustness, and ease of fabrication. The optical system relays radiation from the SOFIA telescope to the detectors, and conditions and spectrally filters the signal radiation, while rejecting as much "noise" radiation as possible. We mounted all optical components on an optical bench cooled to 1.3 K to provide alignment stability, and reduce the difficulty of eliminating stray radiation from warmer components, a difficulty with sensitive long-wavelength systems. The instrument configuration separates the high background part of the volume from the low background areas. The design meets all of the SOFIA telescope interface requirements with no exceptions. Fig. 4 shows a schematic view of the two-stage collimator/camera design, which provides a simple and flexible layout, with enough aberration control to meet image quality goals.

There are actually three collimator/camera relays (as shown in Fig. 5): a first stage that is common to all optical configurations, and two second stages for the separate detector arrays. All use off-axis conics and achieve aberration control by adjusting the conic constants and off-axis decentrations.

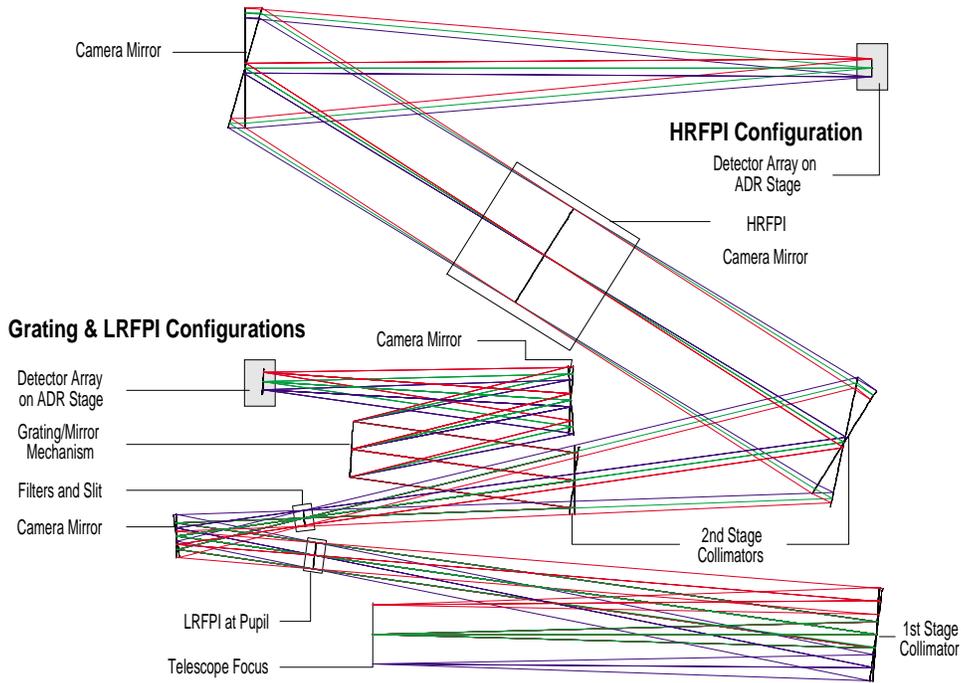

**Figure 4.** Optical Block Diagram - Optical configuration maximizes the use of common optical path and elements to minimize volume and cost.

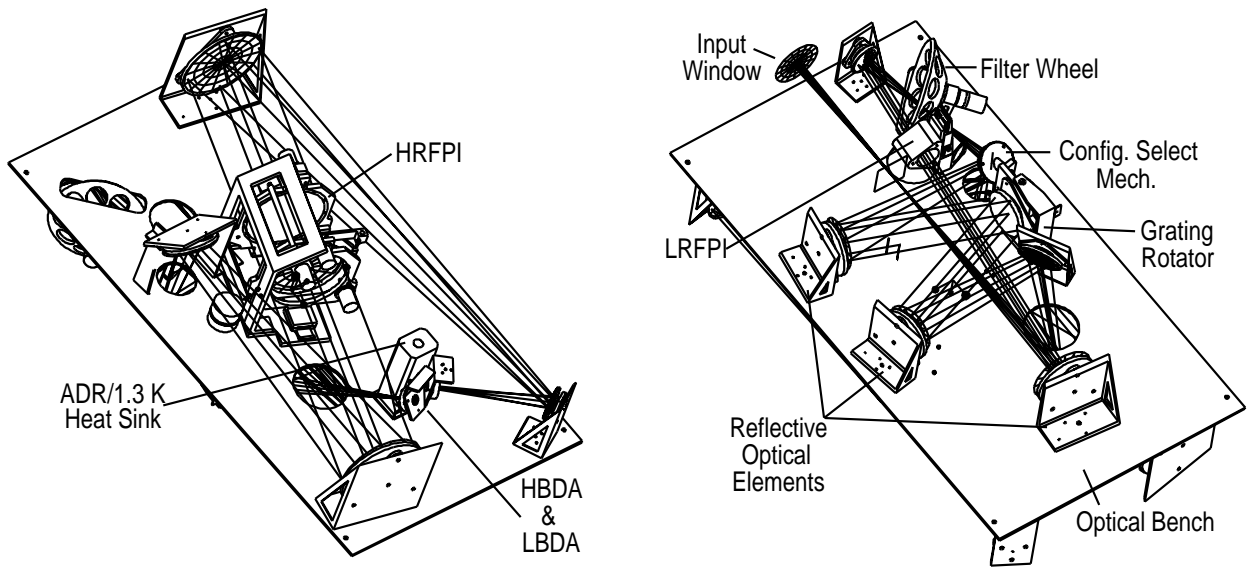

**Figure 5.** SAFIRE Top (Left) and Bottom (Right) Deck Optical Layout

The first stage uses a paraboloid collimator and an ellipsoid camera mirror. The current opto-mechanical layout requires that the telescope focus be 15 cm away from nominal. This induces some small spherical aberration, but it is compensated by adjusting the conic constant of the first stage camera mirror. This needed adjustment, along with the faster image speed (f/5.7 at the intermediate image), makes the first stage camera mirror the most exotic of the SAFIRE mirrors. However, of the mirrors are feasible using our fabrication and testing facilities.

The second stages use only paraboloids. The Grating/LRFPI-only path is an Ebert-Fastie arrangement that controls coma and astigmatism. The HRFPI path uses control of the beam angles on the powered mirrors to effect this. The final images are diffraction limited with ample margin (the geometrical spot radii are all less than 0.1 mm, compared to the 2 mm radius of the first Airy minimum at $\lambda = 250$ μm).

The grating for SAFIRE is placed at a pupil in the second stage of the Grating/LRFPI-only optical path. It operates in 1st and 2nd orders, with spectral resolution of R = 90 to R = 180 over the wavelength range of each order (145 to 300 μm in 2nd order, 300 to 650 μm in 1st order). The spectral image of each order is about three times wider than the detector field of view, so the grating must be scanned to three positions to capture each order completely. This requires that the mechanism have fine rotation capability (30 arc-sec resolution) over a range of ±2°.

The SAFIRE optical elements are each made of a single piece of stress relieved aluminum. This includes the grating, and mirrors that are attached to mechanisms. The optical surfaces will be diamond-turned with integral alignment fiducials and low emissivity coatings. The optical bench will be machined from one large aluminum block. The bench will be pocketed, lightweighted and ribbed. Special attention will be given to stiffness, stability and thermal conduction. The optical bench will slide into the cryostat on rails. The bench will be kinematically clamped and bolted to the cryostat inner shell support rings. Cooling all optical elements to 1.3 K solves the problem of radiation from warm components degrading instrument sensitivity by eliminating any warm components.

Background and scattered light control within the instrument is achieved via close baffling at the telescope and intermediate field images, and at the first and second stage pupil images. Optical-quality surface finishes on the windows and mirrors are available, which should keep scatter to negligible levels within the optical system. Having the detectors on a separate level from the telescope input beam naturally compartmentalizes the system, and additional baffles to segregate different stages of the beam path prevents any contamination of the final image. A final baffle extending from the 0.065 K stage controls the solid viewing angle of the detector array, and provides a long wavelength cut-off filter (1 mm and beyond), to reduce any background from the 1.3 K dewar radiation.

## 3. CRYOGENICS

The cryogenics subsystem maintains the instrument optics and mechanisms at 1.3 K, and the detectors at 0.065 K (see Table 2). The dewar (see Fig. 6) provides the cryogenic environment and the structure required to mount the instrument on the telescope. The dewar vacuum shell has a mounting flange on one end to bolt to the telescope Nasmyth tube. Alignment pins ensure proper alignment with the telescope. A combination of large gussets and isogrid webs on the dewar case provide a rigid support for the instrument. The dewar case is designed to contain the instrument during crash loads. The dewar is sized to fit through the SOFIA aircraft door at no larger than 104 cm wide by 150 cm tall by 200 cm long, while providing the maximum operating volume for the instrument. The dewar weight, including the Adiabatic Demagnetization Refrigerator (ADR), but without optics, mechanisms or attached electronics, is approximately 405 kg.

Table 2. Cryogenics Requirements and robust design implementation enable science operations.

| Derived Requirement | Key Design Parameter |
| --- | --- |
| 48 hour cryogen hold | • 50 liter 4.2 tank; 720 mW heat load budget on 4.2 K |
| | • 10 liter 1.3 K tank; 90 mW heat load budget on 1.3 K |
| 0.065 K detector operating temperature, 24 hours | • ADR; 3.5 μW heat load budget onto ADR |
| Safety | • 1.3K stage burst disk on main shell |
| | • Multiple vent paths out of dewar |
| | • Duct all vents out of aircraft |
| | • FAA approved pressure seal |
| Low thermal input to 4.2 K | • Baffled light path |
| | • Vapor cooled shields |
| | • Kevlar™ suspension system |
| Low thermal input to 1.3 K | • Window on 4.2 K shell |
| | • Low dissipation mechanisms |
| | • Low-power SQUID amplifiers |
| Ease of upgrade/ refurbishment | • Removable detector insert |
| | • Removable dewar covers |
| Allow telescope ±20° rotation | • LHe position and shape |

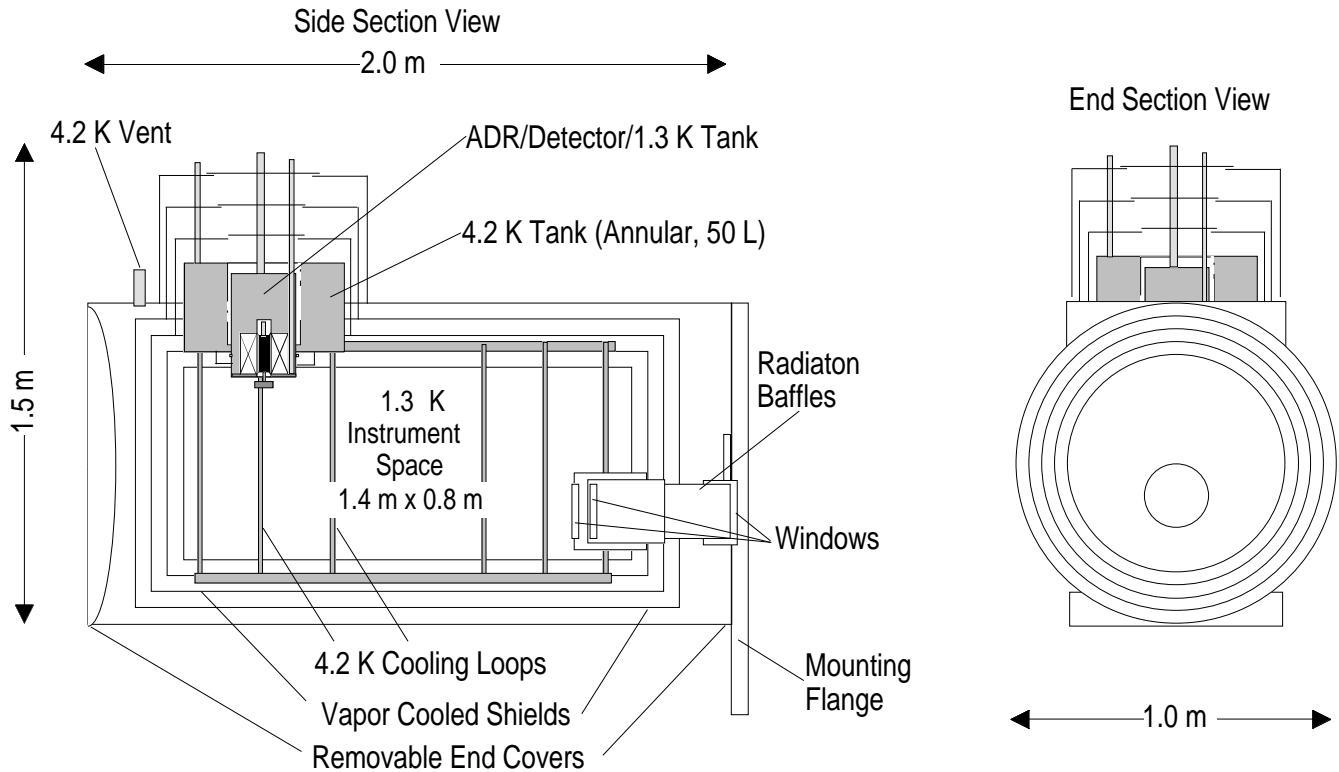

**Figure 6.** Dewar Side and End Views: Modular designs provides easy assembly and access.

The detectors mount to the cold end of the removable cryostat ADR assembly. A kinematic arrangement of Kevlar fibers connects the detectors to the ADR 1.3 K tank. The ADR 1.3 K tank will have a direct mechanical mount to the optical bench. This arrangement provides the metering path between the detectors and the bench and a direct thermal path from the optical bench to the ADR 1.3 K tank.

The SAFIRE detectors must operate with a heat sink temperature at approximately 0.065K to obtain the required sensitivity. An ADR is the most robust and cost-effective way to achieve this temperature. Among the advantages of ADRs is their ability to be operated with simple controls and no moving parts. The baseline ADR design uses the same construction of the salt pill as was used for the XRS ADR[3], which has passed space flight qualification tests (see Fig. 7). We will use some spare parts from XRS, reducing SAFIRE costs. The expected parasitic heat leak of about 3.5 microwatts, again based on results from XRS, will result in a hold time of greater than 24 hours at 0.065 K. This means that no magnetization or heat switch cycling need be done during the mission, delivering a 100% duty cycle. Like the dewar tanks, this salt pill is oversized for the present design, but easily accommodates extra parasitic heat loads from detector wiring and suspension.

The ADR magnet will also use the XRS design, limiting the stray field at the detectors to around 16 gauss at full field, and less than 5% of this at the operating field. The baseline detectors (in particular the SQUID amplifiers) will be magnetically shielded so as to operate well at this field level. The SAFIRE magnet will be cheaper and more robust, by using a higher current to reach 2 Tesla rather than the low current models used by XRS. The magnet is sized to accommodate the large salt pill, so can easily be adapted if a smaller salt pill is used in the future. The higher parasitic heat loads of this type of magnet are about 250 mW and are easily intercepted by the 4.2 K tank.

The baseline heat switch is the XRS gas-gap heat switch[4] which uses a heater-activated getter to turn the conduction on and off. This design is proven but is heavier and has a higher parasitic heat load than promising new techniques of either a $^3$He–$^4$He thermal diode or a capillary flow heat switch. These two new techniques are currently under development at GSFC. If one of these should prove successful, it will be used on SAFIRE.

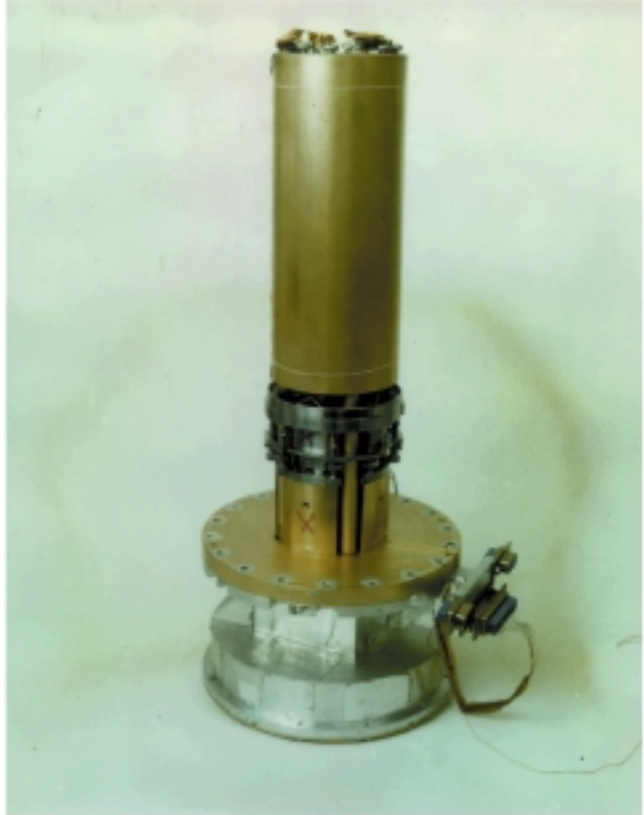

**Figure 7.** Existing XRS ADR Hardware will be used and designs copied to reduce cost.

## 4. DETECTORS

The bolometers we have developed over the last several years can be built into large arrays. Two close-packed, 2-dimensional arrays of silicon microbolometers with frequency independent absorption coatings and having 50% quantum efficiency satisfy the SAFIRE detector requirement for low-background and high-background observational configurations. The arrays are attached to a 0.065 K heat sink provided by the ADR. The baseline array configuration for SAFIRE is two 6x32 element Pop-Up Detector (PUD) arrays utilizing recently established superconducting transition edge sensor (TES) thermometers[2], read out by multiplexed SQUID amplifiers[5]. The SAFIRE instrument is designed ultimately to be fitted with 32x32 arrays of superconducting PUDs, currently under development through a joint NIST / GSFC collaboration which should be completed by 2002 and may be ready for the first SOFIA flight.

Numerous micromachined silicon bolometers have been fabricated at GSFC. Some of these bolometers – using semiconductor thermistors – were used for the GSFC instrument MIRA onboard the Kuiper Airborne Observatory, were used on the Medium Scale Anisotropy Measurement balloon platform, and are presently in use on the SHARC instrument at Caltech Submillimeter Observatory[6]. They are also presently in use in the SPIFI[7] instrument for observations in Antarctica, with an upgrade to a 4x32 TES array currently being assembled. The bolometers are fabricated using conventional, monolithic silicon batch-processes, combined with well-established silicon micromachining techniques developed at GSFC. The superconducting transitions are very stable and can be thermally cycled indefinitely with no degradation in the transition temperature or normal-state resistance. We now have processes for making devices with a final thickness of 1 μm, which have lower thermal conductances and will be improved further by incorporating phonon scattering techniques now under development. Detectors fabricated using the 1 μm technology are measured to have a phonon-mediated thermal conductance of $7\times10^{-12}$ W/K at 350 mK, which translates to a conductance of $3 \times 10^{-13}$ W/K at 0.1 K, and an electrical NEP (noise equivalent power) $<10^{-18}$ W/√Hz at 100 mK. The performance of bolometers is well predicted by using the phonon-noise limit. When using superconducting thermometers, time constant considerations are reduced through electrothermal feedback[8], in which the effective time constant is reduced by orders of magnitude below the thermal time constant.

Assembling a large, filled two-dimensional array of bolometers involves a methodical integration process, shown in Fig. 8. The bolometers are fabricated in 1x32 PUD linear arrays; PUD linear-array-pairs are then inserted into the array mount. The pairing allows bonding wires of one array to be nested into the gap provided by the asymmetry of the neighboring array, so wires will not be sheared off when a linear-array-pair is inserted or removed from the middle of the array. Array-pairs are then connected to the SQUID amplifiers from the back-side of the array. Connecting from the backside allows easy insertion of array-pairs and protects the PUD array elements from damage during electrical wiring. The amplifier wiring is connected via a low-thermal-conductance silicon-bridge chip. Each linear array connects to a single 1x32 SQUID multiplexed amplifier, which multiplexes the signals from 32 detectors into one output. Gain is provided by a SQUID series array amplifier[9]. The PUD array mount is inserted into the ADR dewar and bolted to the pumped He stage. The Kevlar suspended PUD elements are then thermally connected to the ADR with heat straps through the PUD array's mounting clips. To give an impression of the appearance of a final array, Fig. 9 shows the detectors of a 5x32 after folding and stacking, but without the SQUID amplifier boards underneath.

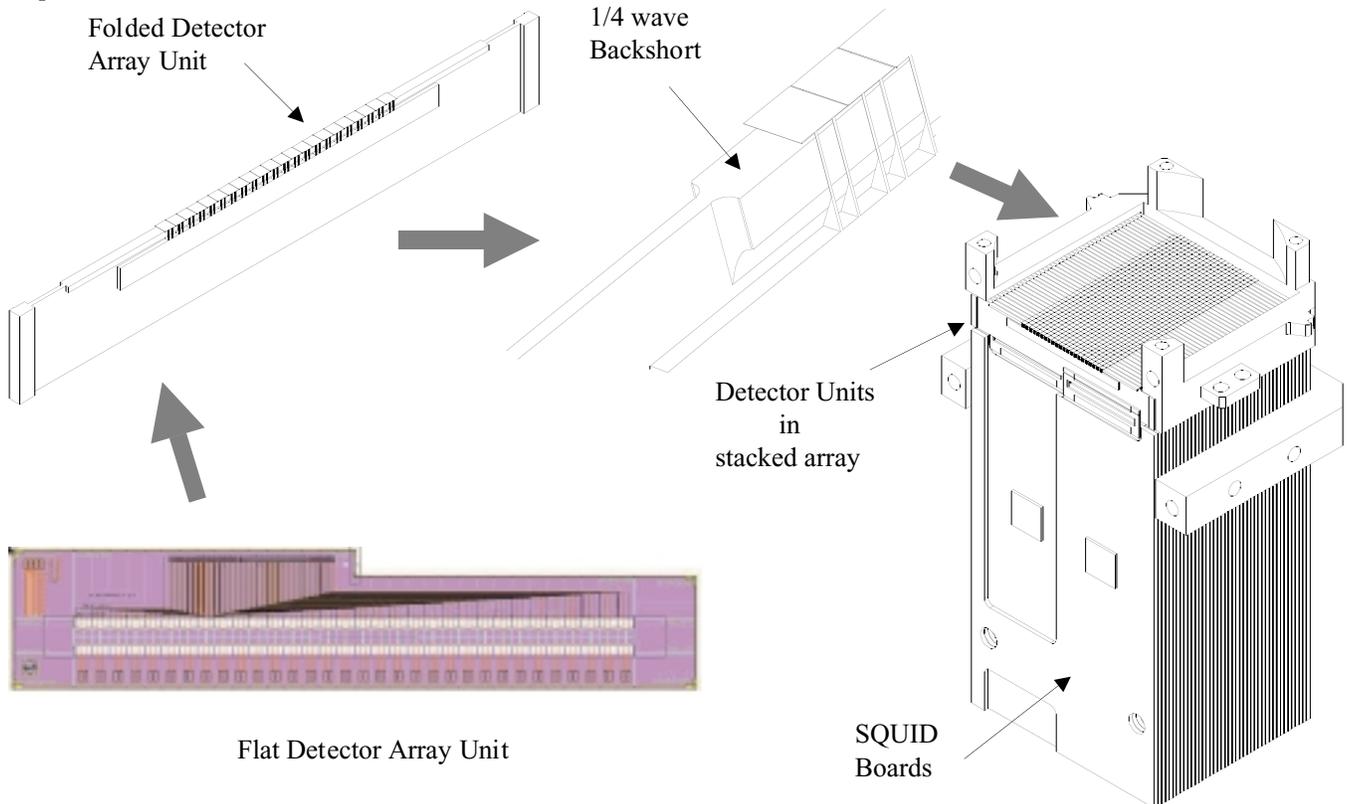

**Figure 8**. Process of assembling two-dimensional PUD arrays. To demonstrate the extensibility of this method, a 32x64 array is shown; the initial SAFIRE array design consists of 6x32 elements.

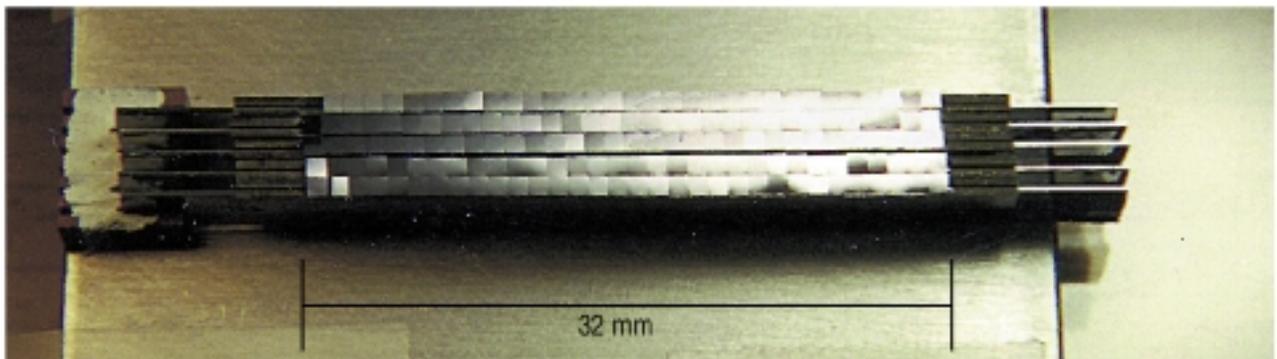

**Figure 9.** Test assembly of a 5x32 array of 1 mm$^2$, 1μm thick PUD mechanical models. The detector model shown is identical in geometry to that whose electrical NEP was measured to be $3 \times 10^{-18}$ W/√Hz at 85mK.

The electronics for reading out a SQUID-multiplexed bolometer array are shown in Fig. 10. For simplicity, only one column (1x32 detectors) is shown; an identical circuit is required for every column. The majority of the circuit in Fig. 10 are the SQUIDs themselves, which are fabricated at NIST on a small silicon chip. The remainder of the electronics are predominantly warm and outside the cryostat. A block diagram of the entire data acquisition system for SAFIRE is shown in Fig. 11, assuming 32x32 arrays. To date, the electronics for a 4x32 array have been fabricated and characterized.

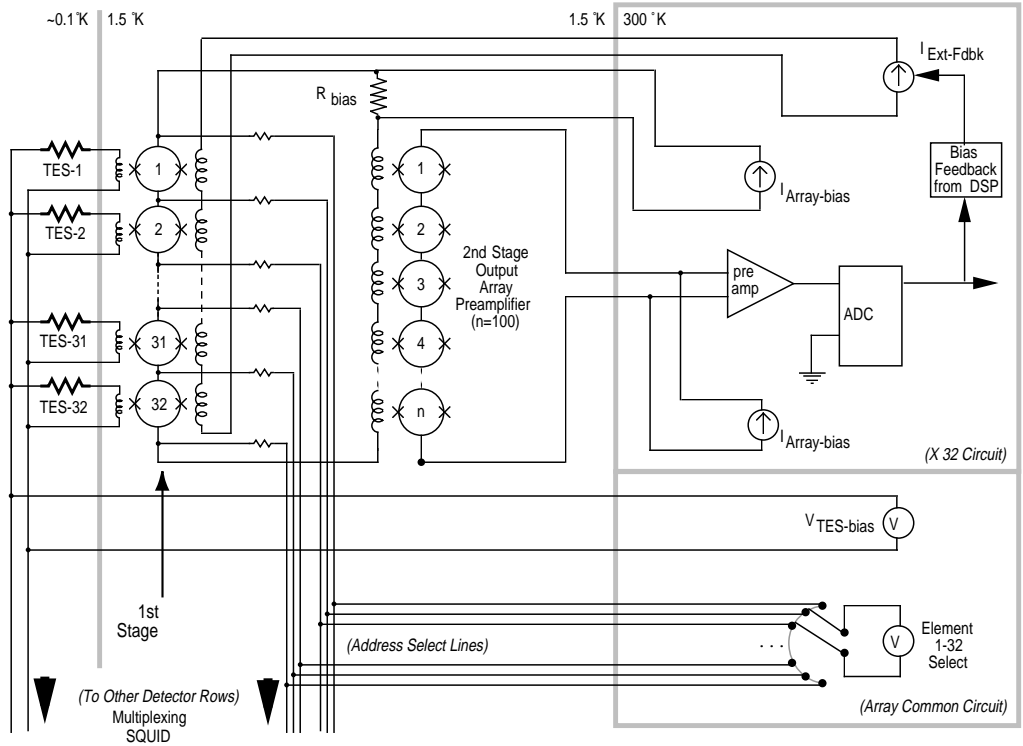

**Figure 10.** Detail of the electronics shows the TESs, the multiplexing SQUIDs, and the SQUID array pre-amplifier.

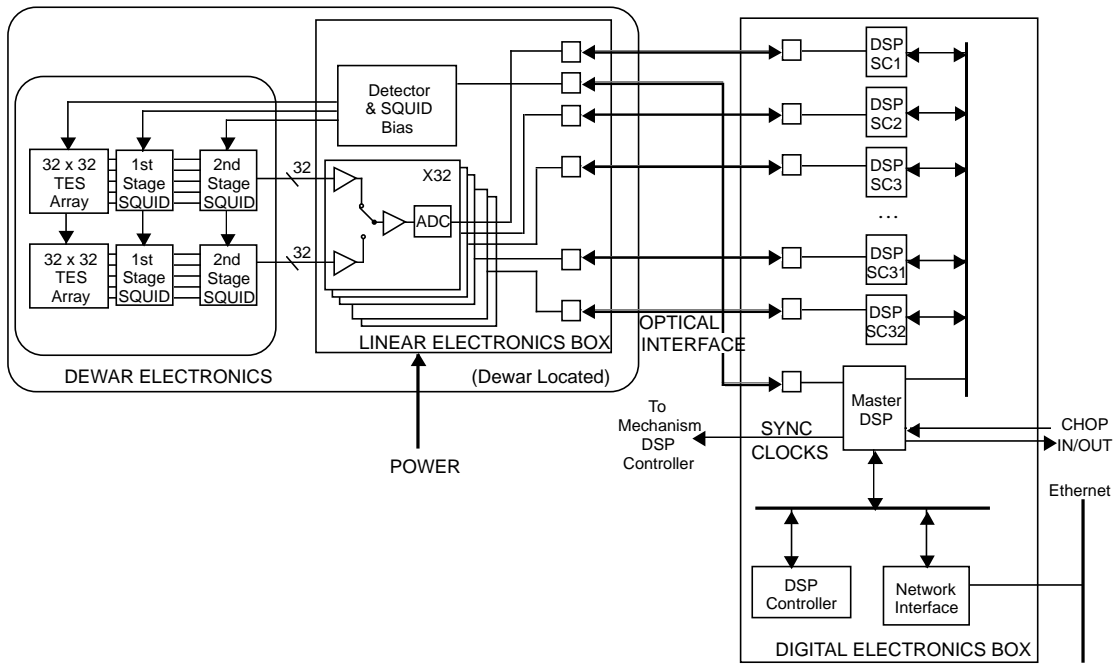

**Figure 11.** Block diagram of the data acquisition system for SAFIRE with two 32x32 arrays.

## 5. CONCLUSION

SAFIRE operates over a broad spectral range which contains many diagnostic lines for studies of energy balance and physical conditions in a wide variety of astrophysical systems. The flexibility to use the instrument to address a wide variety of question is an important characteristic of SAFIRE. Many of the key lines for the possible research projects are inaccessible even from the best ground based sites, making SAFIRE a unique facility for submillimeter astronomy. The combination of the large focal plane arrays and the high sensitivity provided by both the instrument design and the SOFIA telescope creates a capability far beyond that of any existing instrument. SAFIRE will enhance the Space Infrared Telescope Facility (SIRTF) scientific return by providing the ability to follow up SIRTF studies with measurements at longer wavelengths and higher spectral resolution. Given the loss of the $\lambda > 200$ μm capability on SIRTF, SAFIRE provides an important tool for extending observations of interesting objects into the submillimeter as required. SAFIRE, with its spectral resolving power reaching $10^4$, is a powerful complement to future missions such as the Far Infrared and Sub-millimeter Space Telescope (FIRST). The bolometer instrument on FIRST, SPIRE[10], is likely to have a resolving power of ~1000, so for narrow lines on strong continua, SAFIRE can significantly extend investigations initiated on FIRST. SAFIRE can do this because of the large weight and size allowed for instruments on SOFIA as compared to space missions. Also, instruments with more complex mechanisms are acceptable in the SOFIA environment, compared to the less forgiving space environment. FIRST will benefit greatly from SAFIRE's scientific and technological legacy. Collaborators and other users of SAFIRE are actively sought. We are actively soliciting prospective collaborators to generate observation programs to maximize the return from the leap in capabilities offered by SAFIRE. Interested observers are encouraged to register with the SAFIRE observer announcement list at http://pioneer.gsfc.nasa.gov/public/safire.